\documentclass[twocolumn, superscriptaddress, nofootinbib, longbibliography=false]{revtex4-2}
\usepackage[english]{babel}
\usepackage[utf8]{inputenc}
\usepackage{tikz}
\usetikzlibrary{decorations.pathmorphing}
\usepackage{amsmath}
\usepackage{amssymb}
\usepackage[hidelinks]{hyperref}
\usepackage{titlesec}
\usepackage{indentfirst}
\usepackage{dsfont}
\usepackage{graphicx}
\usepackage{pgfplots}
\pgfplotsset{compat=1.18}

\begin{document}

\title{Gravitationally Induced Entanglement Between Particles in Harmonic Traps: Limits for Gaussian States}

\author{Julia Tokarska}
\affiliation{Institute of Theoretical Physics, University of Warsaw, Pasteura 5, 02-093 Warsaw, Poland}
\author{Andrzej Dragan}
\email{dragan@fuw.edu.pl}
\affiliation{Institute of Theoretical Physics, University of Warsaw, Pasteura 5, 02-093 Warsaw, Poland}
\affiliation{Centre for Quantum Technologies, National University of Singapore, 3 Science Drive 2, 117543 Singapore, Singapore}

\begin{abstract}
Gravitationally induced entanglement has been proposed as a probe of the quantum nature of gravity. This work analyzes a system of two particles in harmonic traps interacting only through gravity, considering thermal and two-mode squeezed initial states. For thermal states, a maximum temperature is identified above which entanglement cannot be generated, and for fixed system parameters an optimal trap frequency that maximizes the logarithmic negativity is found. Squeezing the initial state does not further enhance the entanglement generation, but increases the temperature range over which it can be observed. Extending the analysis to general Gaussian states, an upper bound on the achievable entanglement is derived and shown to be saturated, for example, by ground and squeezed states. The results show that the amount of entanglement generated in this setup is extremely small, highlighting the experimental challenges of observing gravitationally induced quantum effects.
\end{abstract}

\maketitle

\section{Introduction}
\label{sec:introduction}
Although much effort has gone into formulating a~quantum theory of gravity, it remains unclear whether gravity is fundamentally quantum, making experimental verification of its nature essential. A natural approach is to study gravitons, which has long been considered beyond reach \cite{Rothman2006-ti}. Recent work, however, has proposed ways to detect individual quanta of gravitational radiation \cite{Tobar2024-mt}, though the results could still be explained by a classical description of gravity \cite{PhysRevD.109.044009}.

An alternative approach is to study systems in which gravity mediates interactions between quantum objects \cite{PhysRevLett.119.240401, PhysRevLett.119.240402}. In such setups, the appearance of entanglement between two systems interacting only via gravity would indicate that it acts as a~quantum mediator, since such correlations cannot arise classically.

While the proposals in \cite{PhysRevLett.119.240401, PhysRevLett.119.240402} focused on massive objects placed in superpositions with large spatial separations, there is growing interest in continuous-variable systems, which are simpler to implement experimentally. Recent studies have investigated gravitational entanglement between particles in harmonic traps prepared in the ground state \cite{PhysRevD.105.106028}, coherent states \cite{PhysRevD.107.106018}, squeezed states \cite{Qvarfort_2020, Krisnanda2020-ia}, or thermal states \cite{Krisnanda2020-ia, Datta_2021}. Other work has explored the possibility of detecting gravity-induced squeezing \cite{Datta_2021}.

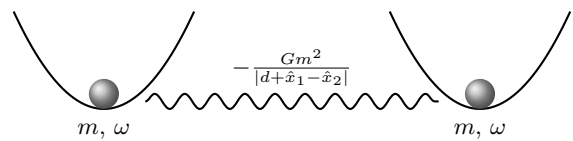
\begin{figure}[t]
	\centering
	\begin{tikzpicture}[scale=1]
		\begin{tikzpicture}[xshift=3.8cm, yshift=0.5cm]
			\def\masssize{0.2}
			\def\parabolaheight{3}
			\def\trapsep{5}
			
			\draw[thick, domain=-1.2:1.2, smooth, variable=\x] 
			plot ({-\trapsep/2 + \x}, {\parabolaheight*0.3*\x*\x});
			
			\draw[thick, domain=-1.2:1.2, smooth, variable=\x] 
			plot ({\trapsep/2 + \x}, {\parabolaheight*0.3*\x*\x});
			
			\shade[ball color=gray!60] (-\trapsep/2, 0.2) circle (\masssize);
			\shade[ball color=gray!60] (\trapsep/2, 0.2) circle (\masssize);
			
			\node at (-\trapsep/2, -0.3) {$m$, $\omega$};
			\node at (\trapsep/2, -0.3) {$m$, $\omega$};
			
			\draw[decorate, decoration={snake, amplitude=1mm, segment length=4mm}, thick] 
			(-\trapsep/2 + \masssize+0.35, 0.1) -- (\trapsep/2 - \masssize-0.35, 0.1);
			
			\node at (0, 0.55) {$-\frac{Gm^2}{|d+\hat{x}_1-\hat{x}_2|}$};
			
		\end{tikzpicture}
	\end{tikzpicture}
	\caption{Experimental setup: two particles of mass $m$ trapped in harmonic potentials with frequency $\omega$, interacting only via gravity.}
	\label{fig:exp_setup}
\end{figure}

This work analyzes a system of two particles of mass $m$, each trapped in a separate harmonic potential with frequency $\omega$ and separated by a distance $d$, interacting only via gravity (Fig. \ref{fig:exp_setup}). The Hamiltonian is
\begin{equation}
	\label{H_full}
	\begin{split}
		\hat{H}=\frac{\hat{p}_1^{\;2}+\hat{p}_2^{\;2}}{2m}+\frac{1}{2}m\omega^2(\hat{x}_1^{\;2}+\hat{x}_2^{\;2})\\-\frac{Gm^2}{|d+\hat{x}_1-\hat{x}_2|},
	\end{split}
\end{equation}
where $\hat{p}_i$ and $\hat{x}_i$ are the momentum and position operators of the $i$th particle. The gravitational interaction, following \cite{PhysRevLett.119.240402}, is approximated as Newtonian.

This work analyzes entanglement generation from thermal and two-mode squeezed states, quantifying the entanglement produced during their evolution. For given values of $d$, $\omega$ and $m$, a temperature is identified above which thermal states remain separable. The analysis further reveals that, for fixed $d$, $m$ and temperature, an optimal trapping frequency exists that maximizes the logarithmic negativity. Extending the study to general Gaussian initial states, an upper bound on entanglement enhancement is established, along with examples of states that saturate this bound.

The paper is organized as follows. Section \ref{sec:evolution} presents the time evolution of the system. Section \ref{sec:thermal} discusses entanglement generated from thermal initial states, while Section \ref{sec:squeezed} focuses on two-mode squeezed states. Section \ref{sec:general} derives an upper bound on entanglement enhancement. Finally, Section \ref{sec:summary} summarizes the main results.

\section{Time evolution}
\label{sec:evolution}

The analysis focuses on masses prepared near their ground state in the harmonic traps, where relative position fluctuations are small compared to the separation $d$. In this regime, the Hamiltonian~(\ref{H_full}) can be approximated as
\begin{equation}
	\begin{split}
		\hat{H}\approx&\; \frac{\hat{p}_1^2+\hat{p}_2^2}{2m} + \frac{1}{2}m\omega'^2\left[\left(\hat{x}_1+\delta\right)^2+\left(\hat{x}_2-\delta\right)^2\right]\\
		&+\frac{2Gm^2}{d^3}\left(\hat{x}_1+\delta\right)\left(\hat{x}_2-\delta\right)+\text{const},
	\end{split}
\end{equation}
with the parameters $\omega'$ and $\delta$ defined as
\begin{equation}
	\omega'^2=\omega^2-\frac{2Gm}{d^3},\quad \delta=\frac{Gmd}{\omega^2d^3-4Gm}.
\end{equation}
and a constant term that can be ignored.

The Hamiltonian is conveniently expressed in terms of the dimensionless operators $\hat{q}_i$ and $\hat{k}_i$, defined as
\begin{equation}
	\begin{split}
	\hat{q}_{1,2}&=\sqrt{\frac{m\omega'}{\hbar}}\;\left(\hat{x}_{1,2}\pm \delta\right),\\ \hat{k}_{1,2}&=\frac{1}{\sqrt{\hbar m\omega'}}\;\hat{p}_{1,2},
	\end{split}
\end{equation}
which satisfy the commutation relations:
\begin{equation}
	[\hat{q}_i,\hat{k}_j]=i\delta_{ij}.
\end{equation}
In this basis, the Hamiltonian takes the form:
\begin{equation}
	\label{H_qp}
	\begin{split}
	\hat{H}=\frac{1}{2}\hbar\omega'(\hat{q}_1^2+\hat{q}_2^2+\hat{k}_1^2+\hat{k}_2^2)\\+\frac{Gm \hbar }{d^3\omega'}(\hat{q}_1\hat{q}_2+\hat{q}_2\hat{q}_1).
	\end{split}
\end{equation}
This expression can be written more compactly by introducing the vector $\hat{\boldsymbol{X}}$ and the matrix $\boldsymbol{H}$:
\begin{equation}
	\hat{\boldsymbol{X}}=\begin{bmatrix}
		\hat{q}_1  \\ \hat{k}_1 \\ \hat{q}_2 \\ \hat{k}_2
	\end{bmatrix}, \quad
	\boldsymbol{H}=\frac{1}{2}\hbar \omega' \begin{bmatrix}
		1 & 0 & \varepsilon & 0\\
		0 & 1 & 0 & 0\\
		\varepsilon & 0 & 1 & 0\\
		0 & 0 & 0 & 1
	\end{bmatrix},
\end{equation}
where $\varepsilon=\frac{2Gm}{d^3\omega'^2}$. In this notation, the commutation relations are given by:
\begin{equation}
	[\hat{X}_j,\hat{X}_k]=i\Omega_{jk},
\end{equation}
where
\begin{equation}
	\boldsymbol{\Omega}=\bigoplus_{i=1}^N \boldsymbol{\omega},\quad 
	\boldsymbol{\omega}=\begin{bmatrix}
		0 & 1\\ -1 & 0
	\end{bmatrix},
\end{equation}
and the Hamiltonian then takes the form:
\begin{equation}
	\label{H_XMX}
	\hat{H}=\hat{\boldsymbol{X}}^T\boldsymbol{H}\hat{\boldsymbol{X}}.
\end{equation}

The initial states considered here are Gaussian, and since the Hamiltonian is quadratic in the operators $\hat{q}_i$ and $\hat{k}_i$, they remain Gaussian under time evolution. Such states are fully described by their vector of first moments and the covariance matrix of the $\hat{q}_i$ and $\hat{k}_i$ operators. The time evolution of the covariance matrix is given by \cite{doi:10.1142/S1230161214400010}:
\begin{equation}
	\label{sigma_ev}
	\boldsymbol{\sigma}(t)=\boldsymbol{S}\,\boldsymbol{\sigma}(0)\,\boldsymbol{S}^T,
\end{equation}
where\footnote{In \cite{doi:10.1142/S1230161214400010} the symplectic matrix is obtained as $\boldsymbol{S} = e^{\frac{t}{\hbar}\boldsymbol{\Omega H}}$ (Eq.~(76), p.~22); however, the calculation yields an additional factor of $2$ in the exponent.}
$\boldsymbol{S} = e^{\frac{2t}{\hbar}\boldsymbol{\Omega H}}$. For the Hamiltonian~(\ref{H_XMX}), the corresponding symplectic matrix takes the form
\begin{equation}
	\boldsymbol{S}_G=\exp\left(\omega' t\begin{bmatrix}
		0&1&0&0\\
		-1&0&-\varepsilon&0\\\
		0&0&0&1\\
		-\varepsilon&0&-1&0\\
	\end{bmatrix}\right)\\.
\end{equation}
Its elements can be expressed explicitly as combinations of trigonometric functions; however, due to their complexity, the full expressions are omitted here.

Entanglement is quantified using the negativity and the logarithmic negativity, which are well suited for bipartite Gaussian states \cite{Horodecki2009-wy}:
\begin{equation}
	\mathcal{N}(\rho)=\frac{\|\tilde{\rho}\|-1}{2},
	\qquad
	E_{\mathcal{N}}=\log_2\|\tilde{\rho}\|.
\end{equation}
Here $\rho$ denotes the density matrix of the bipartite system, $\tilde{\rho}$ its partial transpose, and $\|A\|=\mathrm{Tr}\sqrt{A^{\dagger}A}$ the trace norm. For bipartite Gaussian states, positivity of the partially transposed density matrix is equivalent to separability, ensuring that these measures detect all entangled states \cite{PhysRevLett.84.2726}.

For a bipartite Gaussian state with covariance matrix
\begin{equation}
	\boldsymbol{\sigma}=
	\begin{bmatrix}
		\boldsymbol{\alpha} & \boldsymbol{\gamma} \\
		\boldsymbol{\gamma}^T & \boldsymbol{\beta}
	\end{bmatrix},
\end{equation}
where $\boldsymbol{\alpha}$, $\boldsymbol{\beta}$, and $\boldsymbol{\gamma}$ are $2\times2$ matrices, the entanglement is determined by the symplectic eigenvalue \cite{PhysRevA.72.032334}
\begin{equation}
	\label{ni}
	\nu = \sqrt{2\left(\Delta - \sqrt{\Delta^2 - 4\det\boldsymbol{\sigma}}\right)},
\end{equation}
with $\Delta = \det\boldsymbol{\alpha} + \det\boldsymbol{\beta} - 2\det\boldsymbol{\gamma}$. The negativity and logarithmic negativity are then given by \cite{PhysRevA.72.032334}:
\begin{equation}
	\begin{split}
		\mathcal{N} &= \max\left(0, \frac{1-\nu}{2\nu}\right), \\
		E_{\mathcal{N}} &= \max\left(0, -\log_2 \nu\right).
	\end{split}
\end{equation}
The state is entangled if $\nu < 1$.

\section{Thermal states}
\label{sec:thermal}

We start by examining thermal states, which naturally arise in realistic experimental settings where perfect ground-state preparation cannot be achieved. A thermal state at temperature $T$ is described by the density matrix
\begin{equation}
	\begin{split}
		\rho(T) &= N \sum_{n,m=0}^{\infty} 
		e^{-\frac{\hbar \omega}{k_B T}(n+m)} 
		\lvert n,m\rangle \langle n,m\rvert, \\
		N &= \left(1-e^{-\frac{\hbar \omega}{k_B T}}\right)^2 ,
	\end{split}
\end{equation}
where $\lvert n,m\rangle$ denote the eigenstates of the uncoupled Hamiltonian
\begin{equation}
	\hat{H}_\omega = \hat{H}_1 + \hat{H}_2, 
	\qquad 
	\hat{H}_i = \frac{\hat{p}_i^2}{2m} + \frac{1}{2} m \omega^2 \hat{x}_i^2 .
\end{equation}
These eigenstates satisfy
\begin{equation}
	\begin{split}
		\hat{H}_1 \lvert n,m\rangle &= \hbar\omega\left(n+\tfrac{1}{2}\right)\lvert n,m\rangle,\\
		\hat{H}_2 \lvert n,m\rangle &= \hbar\omega\left(m+\tfrac{1}{2}\right)\lvert n,m\rangle .
	\end{split}
\end{equation}

The initial covariance matrix is given by
\begin{equation}
	\boldsymbol{\sigma}_T(0) = \tfrac{1}{2}\,\theta\,\mathds{1},
\end{equation}
where $\theta = \coth\!\left(\frac{\hbar \omega}{2 k_B T}\right)$. After time evolution, the covariance matrix becomes
\begin{equation}
	\boldsymbol{\sigma}_T(t)
	= \boldsymbol{S}_G\,\boldsymbol{\sigma}_T(0)\,\boldsymbol{S}_G^T
	= \tfrac{1}{2}\,\theta\,\boldsymbol{S}_G \boldsymbol{S}_G^T .
\end{equation}
Evaluating Eq.~(\ref{ni}) with this covariance matrix provides an expression for the generated entanglement. To extract the dominant physical contribution, we expand the result to first order in $\varepsilon$. Under the assumption $\omega'^2 \simeq \omega^2$, this leads to\footnote{The treatment of thermal states in \cite{Krisnanda2020-ia} produces results compatible with Eq.~(\ref{ni_term_approx}). The maximum temperature allowing for entanglement was previously determined in \cite{Datta_2021}, which examined both gravitationally induced entanglement and squeezing. In contrast, the present analysis further reveals the presence of an optimal trap frequency for fixed mass and temperature, a feature that has not been explicitly discussed in the prior literature.}:
\begin{equation}
	\label{ni_term_approx}
	\nu_T(t) \approx {\theta}\left(1-\frac{2Gm}{d^3\omega^2}|\sin(\omega t)|\right).
\end{equation}
The resulting state is entangled if $\nu_T < 1$. This condition defines a threshold temperature above which the system remains separable:
\begin{equation}
	\label{T_max}
	T_{\mathrm{max}}\approx\frac{\hbar \omega}{k_B \log\left(\frac{d^3\omega^2}{Gm}\right)}
\end{equation}
and is governed predominantly by the trap frequency $\omega$, showing an almost linear dependence. Its value is extremely small; for example, for $d~=~10^{-4}$~m, $\omega = 1$~kHz, and $m = 10^{-15}$~kg, one finds $T_{\mathrm{max}}~\sim~10^{-10}$~K.

As expected for particles confined in harmonic traps, $\nu_T$ exhibits an oscillatory time dependence. Reducing the trap separation increases the coupling strength, allowing larger amounts of entanglement to build up. This separation, however, cannot be reduced arbitrarily. At~sufficiently short distances, the Casimir-Polder interaction becomes significant and must be taken into account.

In this regime, the system Hamiltonian acquires an additional contribution which, for spherical objects separated by a distance much larger than their size, is given by \cite{PhysRev.73.360}:
\begin{equation}
	\frac{23 \hbar c R^6}{4 \pi D^7}
	\left(\frac{\epsilon-1}{\epsilon+2}\right)^2 ,
\end{equation}
where $c$ is the speed of light, $R$ the sphere radius, $D$ their separation, and $\epsilon$ the dielectric constant of the material. A comparison with the gravitational potential shows that the distance above which the Casimir--Polder contribution is at least an order of magnitude weaker than gravity is of order $10^{-4}\,\mathrm{m}$. This value is therefore adopted as the minimal reasonable separation $d$.

At shorter distances, Casimir--Polder forces become comparable to or dominate over gravity, leading to entanglement that is no longer of gravitational origin. For~this reason, the following calculations use $d = 10^{-4}\,\mathrm{m}$. It is worth noting that several proposals \cite{PhysRevA.102.062807, PhysRevResearch.6.013199, PhysRevResearch.5.043170} suggest placing a conducting plate between the masses to suppress electromagnetic interactions, thereby allowing the traps to be positioned closer together.

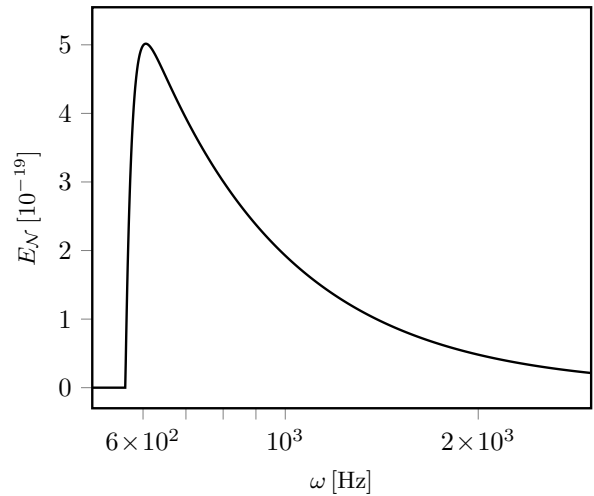
\begin{figure}[t]
	\centering
	\begin{tikzpicture}
	\begin{axis}[
		width=0.95\linewidth,
		height=0.8\linewidth,
		xmode=log,
		xmin=5e2,
		xmax=3e3,
		ymin=-0.3,
		xlabel={$\omega\,[\mathrm{Hz}]$},
		ylabel={$E_{\mathcal N}\,[10^{-19}]$},
		ticklabel style={font=\normalsize},
		label style={font=\normalsize},
		xtick={600,700,800,900,1000,2000}, xticklabels={$6\!\times\!10^{2}$,,,,$10^{3}$,$2\!\times\!10^{3}$},
		ytick={0,1,2,3,4,5},
		grid=none,
		axis lines=box,
		xtick pos=left,
		ytick pos=left,
		tick align=outside,
		line width=0.9pt,
		]
		\addplot[black]
		table[x=omega, y=ENscaled] {EN_vs_omega.dat};
	\end{axis}
    \end{tikzpicture}
	\caption{$E_{\mathcal{N}}$ as a function of $\omega$ for $m=10^{-15}$ kg and $T=10^{-10}$ K.}
	\label{fig:EN_w}
\end{figure}

The dependence of $\nu_T$ on the trapping frequency $\omega$ is now examined for fixed temperature $T$ and mass $m$ (Fig.~\ref{fig:EN_w} shows the logarithmic negativity $E_{\mathcal{N}}$ as a function of $\omega$ for $T=10^{-10}\,\mathrm{K}$ and $m=10^{-15}\,\mathrm{kg}$). For~small values of $\omega$, while still sufficiently large for $\varepsilon$ to remain a~small parameter so that Eq.~(\ref{ni_term_approx}) provides a~valid approximation, the parameter $\theta$ is large, resulting in $\nu_T > 1$ and hence $E_{\mathcal{N}} = 0$. As $\omega$ increases, $\theta$ approaches unity, and above a threshold frequency $\omega_{\min}$ the condition $\nu_T < 1$ is satisfied, indicating the onset of entanglement. For~larger $\omega$, the parameter $\varepsilon$ decreases, causing $\nu_T$ to increase again. In the limit $\omega \to \infty$, one has $\theta \to 1$ and $\varepsilon \to 0$, which implies $\nu_T \to 1$ and consequently $E_{\mathcal{N}} \to 0$. Therefore, for a given temperature and mass, an optimal trapping frequency $\omega_{\mathrm{opt}}$ exists that maximizes the logarithmic negativity.

Numerical calculations for temperatures in the range $10^{-15}\,\mathrm{K}$ to $1\,\mathrm{K}$ and mass $m = 10^{-15}\,\mathrm{kg}$ show that $\omega_{\min}$ is~only slightly smaller than $\omega_{\mathrm{opt}}$. On a~log--log scale, both frequencies exhibit an approximately linear dependence on temperature. In particular, $\omega_{\mathrm{opt}}$ is well described by
\begin{equation}
	\omega_{\mathrm{opt}} \approx 1.4 \times 10^{13}\, T^{1.04},
\end{equation}
with $\omega_{\mathrm{opt}}$ expressed in hertz and $T$ in kelvin. The dependence on the mass is weak over the physically relevant range.

Current experiments can measure values of $\nu$ close to unity with an uncertainty of approximately $0.05$ \cite{doi:10.1126/science.abf2998}. At~this level of precision, a clear demonstration of entanglement therefore requires $\nu_T < 0.95$. Achieving this regime demands both very large masses and extremely low temperatures, which are currently beyond experimental reach. While temperatures as low as $10^{-11}\,\mathrm{K}$ have been achieved, the corresponding masses (Bose--Einstein condensates) are of order $10^{-21}\,\mathrm{kg}$ \cite{PhysRevLett.127.100401}, yielding $\varepsilon \approx 10^{-19}\; \mathrm{Hz}^2/\omega^2$. Conversely, more massive systems, such as microspheres with masses around $10^{-9}\,\mathrm{kg}$, cannot be cooled to such temperatures with present techniques \cite{PhysRevLett.131.043603} and have diameters of order $10^{-4}\,\mathrm{m}$. To suppress interactions other than gravity, the separation between such objects would therefore need to be increased, which in turn leads to larger values of $\nu_T$.

\section{Two-mode squeezed states}
\label{sec:squeezed}

Let us now examine the evolution of two-mode squeezed states, motivated by their ability to amplify weak physical effects. A well-known example is their use in LIGO detectors, where squeezing has significantly improved measurement sensitivity \cite{Aasi2013-bm}. This suggests that gravitationally induced entanglement may be enhanced when the initial state is already squeezed, potentially exceeding that generated from initially separable states.

The initial covariance matrix is given by\footnote{This form differs from those considered in \cite{Qvarfort_2020, Krisnanda2020-ia}, where single-mode squeezing applied independently to each particle was analyzed.}
\begin{equation}
	\boldsymbol{\sigma}_s(0) = \frac{1}{2}
	\begin{bmatrix}
		\cosh r & 0 & \sinh r & 0 \\
		0 & \cosh r & 0 & -\sinh r \\
		\sinh r & 0 & \cosh r & 0 \\
		0 & -\sinh r & 0 & \cosh r
	\end{bmatrix},
\end{equation}
where $r$ is the squeezing parameter.

From Eq.~(\ref{ni}), the initial value of $\nu_s$ is
\begin{equation}
	\nu_s(0) = e^{-|r|}.
\end{equation}
Under time evolution, the covariance matrix takes the form
\begin{equation}
	\boldsymbol{\sigma}_s(t) = \boldsymbol{S}_G \boldsymbol{\sigma}_s(0) \boldsymbol{S}_G^{T}.
\end{equation}
The time dependence of $\nu_s(t)$ is obtained by expanding the parameter $\Delta$ to second order in $\varepsilon$:
\begin{equation}
	\label{delta_s}
	\begin{split}
		&\Delta_s \approx \frac{1}{2} \cosh (2 r) + \varepsilon \sinh (2 r) \sin ^2(\omega' t) \\
		&+ \frac{1}{16} \varepsilon^2 \bigl( \cosh (2 r) (-8 \omega'^2 t^2 - 8 \cos (2 \omega' t) + 7) \\
		&+ 2 \sinh^2(r) \cos (4 \omega' t) + 8 \omega'^2 t^2 + 1 \bigr).
	\end{split}
\end{equation}
This expansion yields a first-order expression for $\nu_s(t)$ in $\varepsilon$. For  $r \neq 0$ it takes the form
\begin{equation}
	\label{nu_sq}
	\nu_s(t)=e^{-|r|}\left(1-\frac{r}{|r|}\frac{2Gm}{d^3\omega^2}\sin^2(\omega t)\right).
\end{equation}
This result shows that for $r>0$ the entanglement after time evolution exceeds that of the initial state. In contrast, for $r = 0$ the linear term in $\varepsilon$ in Eq.~(\ref{delta_s}) vanishes, as it is proportional to $\sinh(2r)$. The resulting expression for $\nu_s(t)$ therefore assumes a different form:
\begin{equation}
	\label{nu_r0}
	\nu_s(t)=1-\frac{2Gm}{d^3\omega^2}|\sin(\omega t)|.
\end{equation}
The system is initially in the ground state for both $T~=~0$ ($\theta = 1$) and $r = 0$, so the time dependence of $\nu(t)$ is expected to be the same in these two cases. Indeed, Eq.~(\ref{ni_term_approx}) with $\theta = 1$ reduces to Eq.~(\ref{nu_r0}).

For positive values of $r$, the initial states have a~reduced variance in the difference of particle positions, $\mathrm{Var}(\hat{q}_1-\hat{q}_2) = e^{-r}$. Strongly squeezed states therefore resemble particles in traps with a high frequency $\omega$, which suppresses fluctuations in the position difference and makes entanglement generation more difficult, since it requires a noticeable superposition of the distance between the particles. If the variance becomes too small, the gravitational term in the Hamiltonian (\ref{H_full}) is effectively constant and does not contribute to the dynamics. One might expect that using squeezed states with an increased variance in the position difference (negative~ $r$) would enhance entanglement. However, Eq.~(\ref{nu_sq}) shows that entanglement is actually reduced in this case.

Even though the relative change $\frac{\nu_s(t)-\nu_s(0)}{\nu_s(0)}$ cannot exceed that obtained for thermal states, two-mode squeezing offers an important advantage when finite temperature is taken into account. Compared to the bound (\ref{T_max}), the maximum temperature above which entanglement cannot be observed is significantly increased. A~straightforward calculation gives
\begin{equation}
	T_{\mathrm{max,s}}=\frac{\hbar \omega}{k_B \log\!\left(\frac{e^r+1-\varepsilon}{e^r-1+\varepsilon}\right)} .
\end{equation}
For sufficiently large squeezing, $T_{\mathrm{max,s}}$ can be made arbitrarily large. This suggests that, if achieving sufficiently low temperatures is experimentally challenging, an alternative route toward observing gravitationally induced entanglement is to increase the squeezing of the initial state, or to combine both strategies.

\section{General states}
\label{sec:general}

This section examines how the choice of initial Gaussian state affects entanglement enhancement, with the aim of determining whether any state can produce a~stronger enhancement than those considered above.

Consider a general bipartite covariance matrix written in standard form:
\begin{equation}
	\label{sf}
	\sigma_{\text{sf}} = 
	\begin{bmatrix}
		a & 0 & c & 0 \\
		0 & a & 0 & d \\
		c & 0 & b & 0 \\
		0 & d & 0 & b
	\end{bmatrix}.
\end{equation}
Any bipartite covariance matrix can be transformed into standard form by local symplectic operations~\cite{PhysRevLett.84.2726}, and every standard-form covariance matrix can be prepared from thermal states using two beam splitters and two squeezers~\cite{sym14071485}. Therefore the standard form is not only general, but also corresponds to states that are straightforward to realize experimentally.

The standard form provides a convenient parametrization for general Gaussian states. Expressing the covariance matrix in this way allows derivation of constraints on the maximal entanglement that can be generated via the gravitational interaction for arbitrary initial states.

To guarantee that the covariance matrix represents a~physically valid quantum state, it must satisfy the bona fide condition \cite{doi:10.1142/S1230161214400010}:
\begin{equation}
	\boldsymbol{\sigma} + \frac{i}{2}\boldsymbol{\Omega} \geq 0,
\end{equation}
which imposes the following constraints:
\begin{equation}
	\label{1/2}
	a,b\geq \frac{1}{2},
\end{equation}
\begin{equation}
	\label{square}
	ab\geq c^2+\frac{b}{4a},
\end{equation}
\begin{equation}
	\label{squares}
	2ab-\frac{1}{2}\geq c^2+d^2,
\end{equation}
\begin{equation}
	\label{ech}
	\left(2cd-\frac{1}{2}\right)^2\geq a^2+b^2-4ab(ab-c^2-d^2).
\end{equation}

For the covariance matrix (\ref{sf}) evolved according to Eq.~(\ref{sigma_ev}), the parameter $\Delta$ can be expressed as
\begin{equation}
	\Delta_{\text{sf}} \approx \Delta_0 + \alpha \varepsilon + \beta \varepsilon^2,
\end{equation}
where
\begin{equation}
	\begin{split}
		\Delta_0=\;&a^2+b^2-2cd\\
		\alpha=\;&2(a+b)(c-d)\sin^2(\omega t)\\
		\beta=\;&\frac{1}{8}(4(a+b)^2-c^2-d^2-14cd\\
		&-8((a-b)^2+(c-d)^2)\omega^2t^2\\
		&-4((a+b)^2-4cd)\cos(2\omega t)\\
		&+(c-d)((c-d)\cos(4\omega t)\\
		&-8(c+d)\omega t \sin(2\omega t))).
	\end{split}
\end{equation}

Consider the relative change of $\nu$ over time, defined as $\frac{\nu(t)-\nu(0)}{\nu(0)}$. Three scenarios can lead to a relative change that is at least linear in $\varepsilon$: (1) $\Delta$ contains linear terms in $\varepsilon$ and the relative change is proportional to $\sqrt{\varepsilon}$, which will be shown to be impossible for any physical state; (2)~$\Delta$~contains linear terms and the relative change is proportional to $\varepsilon$, for which it will be demonstrated that no enhancement beyond two-mode squeezed states is possible; or (3) $\Delta$ contains no linear terms, yet $\nu$ still includes linear contributions, in which case the behavior is identical to that observed for thermal states.

To begin the analysis, consider the first scenario, where $\alpha \neq 0$ and
\begin{equation}
	\begin{split}
		\nu_g &\approx \Bigl[ 2 \Bigl( 
		\Delta_0 + \alpha \varepsilon \\
		&-\sqrt{\Delta_0^2 + 2 \alpha \Delta_0 \varepsilon - 4 \det \boldsymbol{\sigma}_{\text{sf}}} 
		\Bigr) \Bigr]^{1/2}.
	\end{split}
\end{equation}
If $\Delta_0^2 = 4\det\boldsymbol{\sigma}_{\text{sf}}$, this simplifies to
\begin{equation}
	\nu_g \approx \sqrt{2\Delta_0} \left( 1 - \sqrt{\frac{\alpha \varepsilon}{2 \Delta_0}} \right).
\end{equation}
In this case, the oscillations would scale as $\sqrt{\varepsilon}$, producing a much larger effect, since $\varepsilon$ is very small. However, the conditions $\alpha \neq 0$ and $\Delta_0^2 = 4\det\boldsymbol{\sigma}_{\text{sf}}$ cannot be satisfied simultaneously for any physical covariance matrix.

The equality $\Delta_0^2 = 4\det\boldsymbol{\sigma}_{\text{sf}}$ can be rewritten as~a~quadratic equation for $d$:
\begin{equation}
	\label{d}
	\begin{split}
		0 =\;& 4ab \, d^2 - 4(a^2 + b^2)c \, d \\ 
		&+ (a^2 - b^2)^2 + 4abc^2.
	\end{split}
\end{equation}
A real solution exists if and only if
\begin{equation}
	(c^2 - ab)(a^2 - b^2)^2 \geq 0.
\end{equation}
From Eq.~(\ref{square}) it follows that this condition holds only when $a = b$. Substituting $a = b$ into Eq.~(\ref{d}) then gives $c = d$, which implies $\alpha = 0$. 

Therefore, the condition $\Delta_0^2 = 4\det\boldsymbol{\sigma}_{\text{sf}}$ leads to $\alpha = 0$, and oscillations proportional to $\sqrt{\varepsilon}$ cannot occur for any physical state.

Next, consider the case $\alpha \neq 0$ and $\Delta_0^2 \neq 4 \det\boldsymbol{\sigma}_{\text{sf}}$. In~this scenario, one finds
\begin{equation}
	\label{bound}
	\begin{split} &\frac{\nu_g(t)-\nu_g(0)}{\nu_g(0)}=\\
		&=-\frac{\alpha\varepsilon}{2\sqrt{\Delta_0^2-4\det\boldsymbol{\sigma}_{\text{sf}}}}\geq -\varepsilon\sin^2(\omega t).
	\end{split}
\end{equation}

The inequality (\ref{bound}) can be rewritten as
\begin{equation}
	(a-b)^2 \bigl( (a+b)^2 - (c+d)^2 \bigr) \geq 0,
\end{equation}
and from Eq.~(\ref{squares}) it follows that
\begin{equation}
	(a+b)^2 - (c+d)^2 \geq a^2 + b^2 - 2cd + \frac{1}{2}.
\end{equation}
Using Eq.~(\ref{ech}) gives
\begin{equation}
	\begin{split}
		&a^2 + b^2 - 2cd + \frac{1}{2} \geq\\
		& a^2 + b^2 - \sqrt{a^2 + b^2 - 4ab(ab - c^2 - d^2)}.
	\end{split}
\end{equation}
It is sufficient to show that
\begin{equation}
	(a^2 + b^2)^2 \geq a^2 + b^2 - 4ab(ab - c^2 - d^2),
\end{equation}
which, after rearranging and applying Eq.~(\ref{squares}) again, reduces to
\begin{equation}
	(a-b)^2 \bigl( (a+b)^2 - 1 \bigr) \geq 0.
\end{equation}
This is satisfied by Eq.~(\ref{1/2}), confirming that inequality (\ref{bound}) holds. Equality is achieved, for example, for two-mode squeezed states with positive $r$.

Finally, consider the case $\alpha = 0$, which implies $c = d$. To obtain oscillations proportional to $\varepsilon$, it is still required that $\Delta_0^2 = 4 \det\boldsymbol{\sigma}_{\rm sf}$; as seen in the first case, this further requires $a = b$. In this situation, $\beta$ simplifies to
\begin{equation}
	\beta = 4(a^2 - c^2) \, \sin^2(\omega t),
\end{equation}
and the corresponding relative change of $\nu_g$ is
\begin{equation}
	\frac{\nu_g(t)-\nu_g(0)}{\nu_g(0)} = -\varepsilon \sqrt{\frac{\beta}{2 \Delta_0}} = -\varepsilon \, |\sin(\omega t)|.
\end{equation}

These results show that, for any physical Gaussian initial state, the relative change of $\nu_g$ is bounded by
\begin{equation}
	\label{limit}
	\frac{\nu_g(t)-\nu_g(0)}{\nu_g(0)} \ge -\varepsilon.
\end{equation}
This provides an upper limit on entanglement enhancement in the system. Thermal and two-mode squeezed states serve as examples of states that saturate this bound, although the overall effect remains very small due to the small value of $\varepsilon$.

\section{Summary}
\label{sec:summary}

We analyzed gravitationally induced entanglement in a~system of two particles confined in harmonic traps and interacting only through gravity. For thermal initial states, entanglement is generated only below a critical temperature given by Eq.~(\ref{T_max}) and exhibits an oscillatory time dependence. Its magnitude increases with mass, while larger separations and higher temperatures suppress it. The dependence on the trap frequency is non-monotonic: for fixed system parameters there exists a minimal trap frequency above which entanglement appears, as well as an optimal frequency at which it is maximized.

Squeezing the initial state does not increase the entanglement generated during evolution compared to the ground state. However, it provides a notable advantage at finite temperature: the range of temperatures over which entanglement generation can be observed is considerably larger. While the maximal increase in entanglement remains limited, initial squeezing allows the effect to appear even if the temperature cannot be lowered to the values required otherwise, offering a potential experimental benefit.

For realistic parameters corresponding to small values of $\varepsilon$, assumed throughout this work, the relative entanglement enhancement is bounded by Eq.~(\ref{limit}) for all Gaussian initial states. This bound is saturated by thermal and squeezed states.

The results indicate that gravitationally induced entanglement in this setup is extremely weak and would require very low temperatures and large masses to be observed experimentally. This suggests that interferometric schemes such as those proposed in Refs.~\cite{PhysRevLett.119.240401,PhysRevLett.119.240402} may offer more favorable prospects for experimental tests.

\bibliography{references}

\end{document}